\documentclass[Physsubmission, Phys]{SciPost}

\hypersetup{
    colorlinks,
    linkcolor={red!50!black},
    citecolor={blue!50!black},
    urlcolor={blue!80!black}
}

\usepackage[bitstream-charter]{mathdesign}
\DeclareSymbolFont{usualmathcal}{OMS}{cmsy}{m}{n}
\DeclareSymbolFontAlphabet{\mathcal}{usualmathcal}
\usepackage{amsmath}
\usepackage[framemethod=TikZ]{mdframed} 
\newenvironment{theo}[1]{%
	\mdfsetup{%
			frametitle={%
				\tikz[baseline=(current bounding box.east),outer sep= 0pt, inner ysep=-.5pt]
				\node[anchor=east,rectangle,fill=blue!40] {\strut #1};}
			}%
	
	\mdfsetup{innertopmargin=2pt,linecolor=blue!40,%
		linewidth=2pt,topline=true,%
		frametitleaboveskip=\dimexpr-\ht\strutbox\relax
	}
	\begin{mdframed}[]\relax%
	}{\end{mdframed}}

\newcommand{\andeq}{\quad \mathrm{and} \quad}

\newcommand{\hbeta}{\beta}
\newcommand{\hgamma}{\gamma}

\begin{document}

\begin{center}{\Large \textbf{
Higher-order $\beta$-functions in the Standard Model and beyond\\
}}\end{center}

\begin{center}
Florian Herren\textsuperscript{1$\star$}
\end{center}

\begin{center}
{\bf 1} Fermi National Accelerator Laboratory, \\[-.3em] Batavia, IL, 60510, USA
\\
* florian.s.herren@gmail.com
\end{center}

\begin{center}
\today
\end{center}


\definecolor{palegray}{gray}{0.95}
\begin{center}
\colorbox{palegray}{
  \begin{tabular}{rr}
  \begin{minipage}{0.1\textwidth}
    \includegraphics[width=35mm]{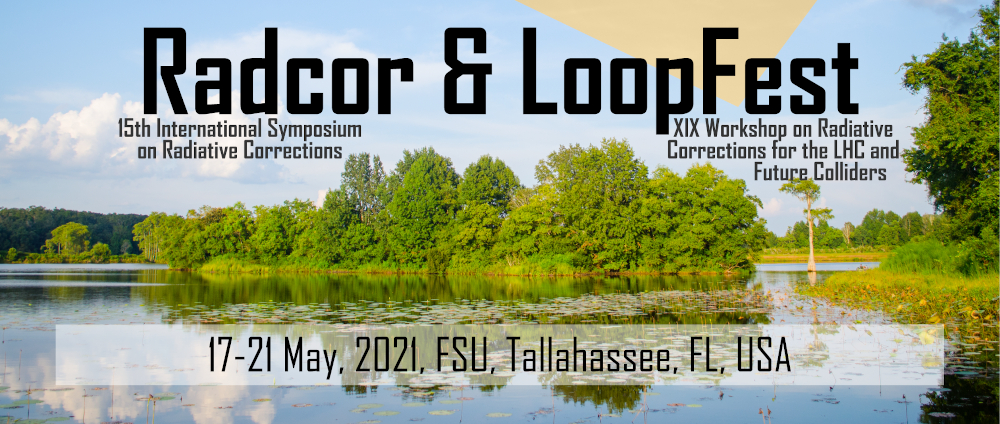}
  \end{minipage}
  &
  \begin{minipage}{0.85\textwidth}
    \begin{center}
    {\it 15th International Symposium on Radiative Corrections: \\Applications of Quantum Field Theory to Phenomenology,}\\
    {\it FSU, Tallahasse, FL, USA, 17-21 May 2021} \\
    \end{center}
  \end{minipage}
\end{tabular}
}
\end{center}

\section*{Abstract}
{\bf
In this contribution we consider the recent computation of the gauge coupling $\beta$-function at four loops and the Yukawa matrix $\beta$-function at three loops in the most
general, renormalizable and four-dimensional quantum field theory. Furthermore, we discuss ambiguities and divergences arising in Yukawa matrix $\beta$-functions.
}

\vspace{10pt}
\noindent\rule{\textwidth}{1pt}
\tableofcontents\thispagestyle{fancy}
\noindent\rule{\textwidth}{1pt}
\vspace{10pt}

\section{Introduction}
$\beta$-functions play a crucial role in phenomenological and theoretical studies of quantum field theories (QFTs), such as the Standard Model of Particle Physics (SM).
They determine the dependence of coupling constants on the renormalization scale $\mu$:
\begin{align}
\mu^2\frac{\mathrm{d}}{\mathrm{d}\mu^2}\frac{\alpha_i}{\pi} = \beta_{\alpha_i}\left(\{\alpha\},\epsilon\right)~.
\end{align}
Here, $\alpha_i$ can be any gauge, Yukawa or quartic scalar coupling, and the respective $\beta$-function can depend on any coupling of the QFT, denoted by the set $\{\alpha\}$.

In non-Abelian gauge theories $\beta$-functions can be negative, thus the gauge coupling vanishes for large scales and as a consequence the
theory is asymptotically free. This property was discovered during the search of a theory of strong interactions \cite{Gross:1973id,Politzer:1973fx} and led
to the establishment of $\mathrm{SU}(3)$ as the gauge group of Quantum Chromodynamics (QCD). Current precision determinations of the strong coupling constant 
require the knowledge of five-loop corrections to the QCD $\beta$-function \cite{Baikov:2016tgj,Herzog:2017ohr,Luthe:2017ttg}.

There are further applications beyond pure QCD, such as the study of gauge coupling unification or the stability of the electroweak vacuum.
To this end, $\beta$-functions for all couplings of the SM or theories beyond it are required. While two-loop $\beta$-functions for the
most general four-dimensional QFT have been known for almost four decades \cite{Machacek:1983tz,Machacek:1983fi,Machacek:1984zw,Jack:1984vj},
higher orders only became available recently. The gauge coupling $\beta$-function for the most general QFT has been computed at three loops \cite{Mihaila:2013dta,Poole:2019kcm}
and this year even the four-loop gauge coupling and three-loop Yukawa $\beta$-functions have become available \cite{Bednyakov:2021qxa,Davies:2021mnc}.

In the following we discuss the computation of the general four-loop gauge coupling and three-loop Yukawa $\beta$-functions and ambiguities as well as divergences
arising in Yukawa $\beta$-functions and give a brief outlook into possible future developments.

\section{$\beta$-functions at 4--3--2-loop order}
In the spirit of the work on the general two-loop $\beta$-functions, Ref.~\cite{Poole:2019kcm} derives a basis of tensor structures for the gauge coupling $\beta$-function
at four loops, the Yukawa $\beta$-function at three loops and combines them with the existing two-loop tensor structures for the quartic coupling $\beta$-function. This
4--3--2-loop ordering is motivated by Osborn's equation \cite{Osborn:1989td,Jack:1990eb,Osborn:1991gm,Jack:2013sha,Baume:2014rla}:\footnote{For the 3--2--1-loop ordering see \cite{Antipin:2013sga}.}
\begin{align}
\partial_I A = T_{IJ}\beta^J~.
\end{align}
Here the derivative w.r.t.~$I$ denotes the derivative w.r.t.~the couplings of the QFT under consideration and $A$ is a scalar function consisting of all
independent contractions of Yukawa matrices, quartic scalar couplings and colour generators, for example:
\begin{align}
\tilde{A} \supset a_{10}^{(3l)} \vcenter{\hbox{\includegraphics[width=1.5cm]{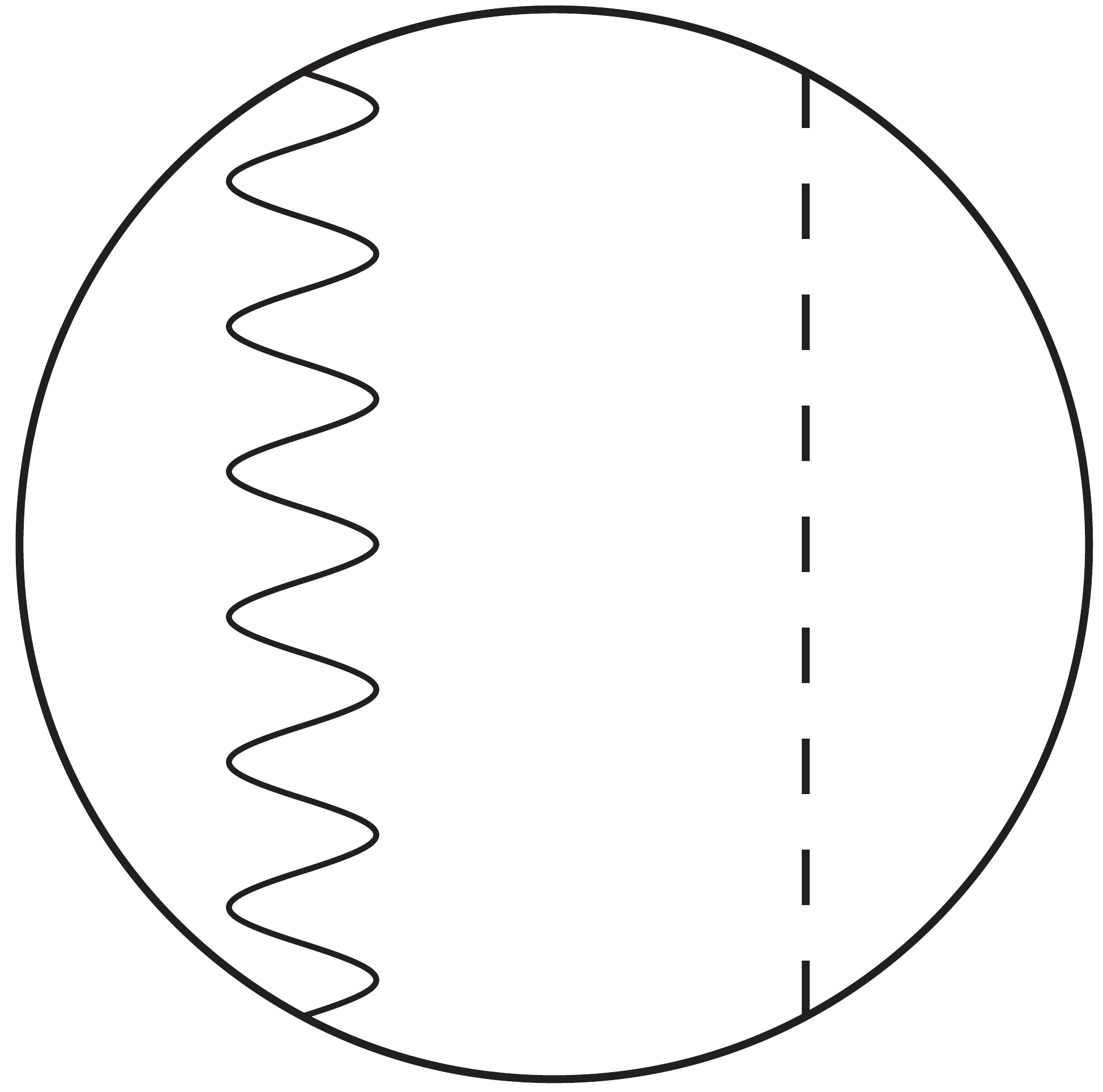}}} + a_{11}^{(3l)} \vcenter{\hbox{\includegraphics[width=1.5cm]{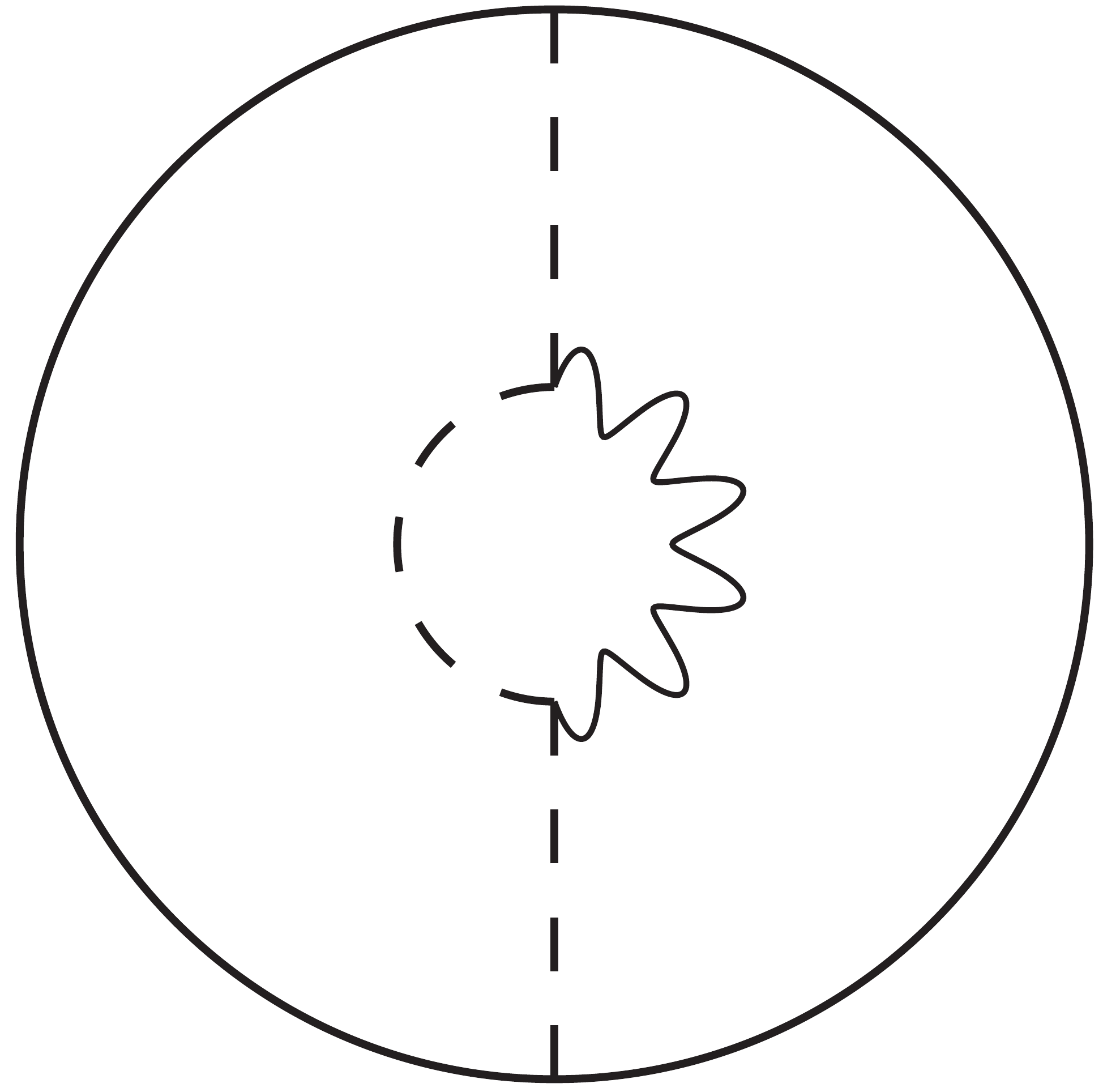}}}~.
\end{align}
Fermion-scalar vertices correspond to Yukawa tensors and gauge-fermion vertices to the gauge generators of the fermions. Gauge couplings themselves are identified with the gauge boson lines.\footnote{For a more detailed
discussion of the diagrammatic notation see \cite{Jack:2014pua}.}
The derivative can be denoted in a pictorial way by removing the corresponding coupling from the tensor structure:
\begin{align}
\partial_I \vcenter{\hbox{\includegraphics[width=1.5cm]{figs/a310-eps-converted-to.pdf}}} = \vcenter{\hbox{\includegraphics[width=1.5cm]{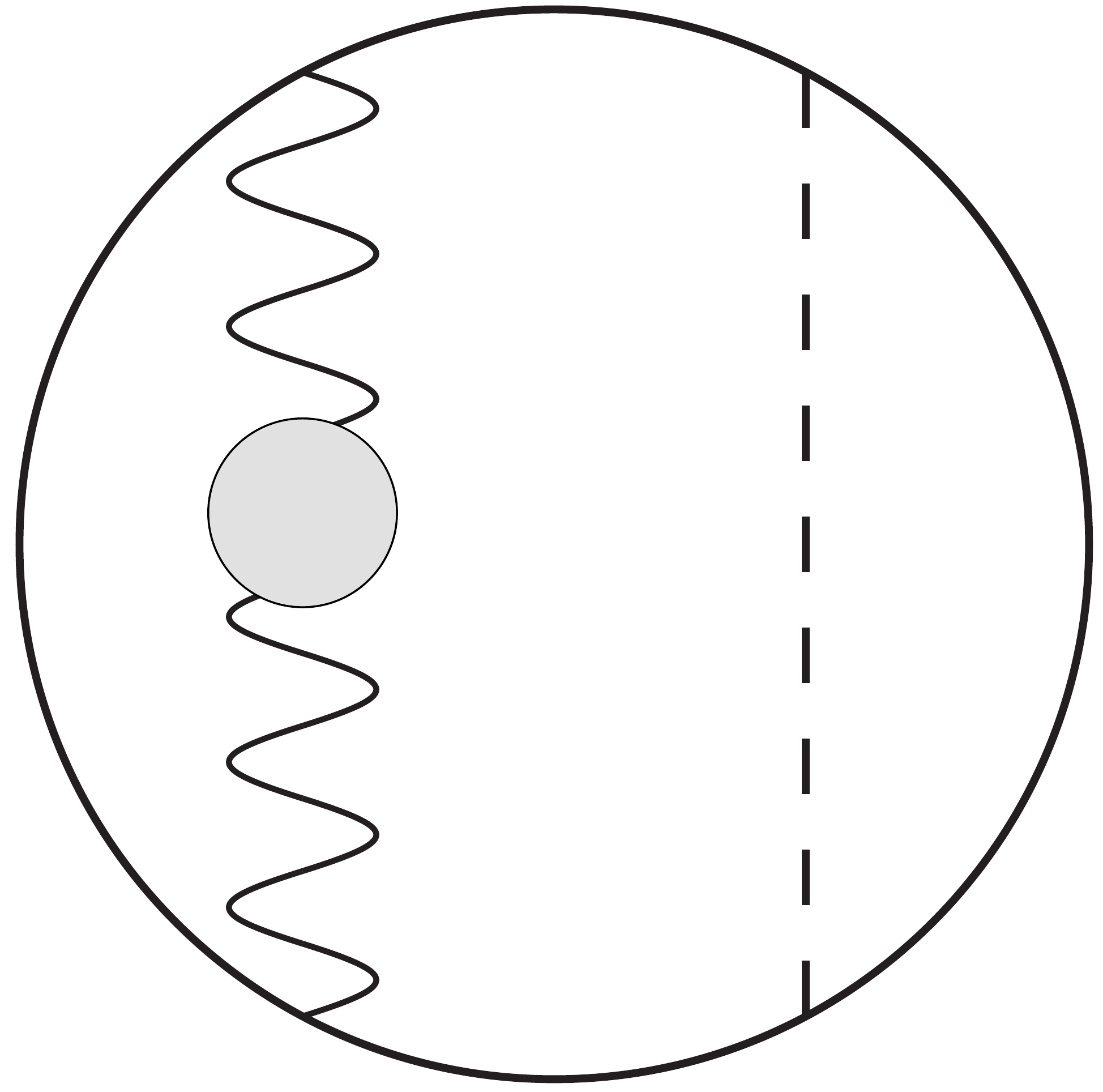}}} + 2\,\vcenter{\hbox{\includegraphics[width=1.5cm]{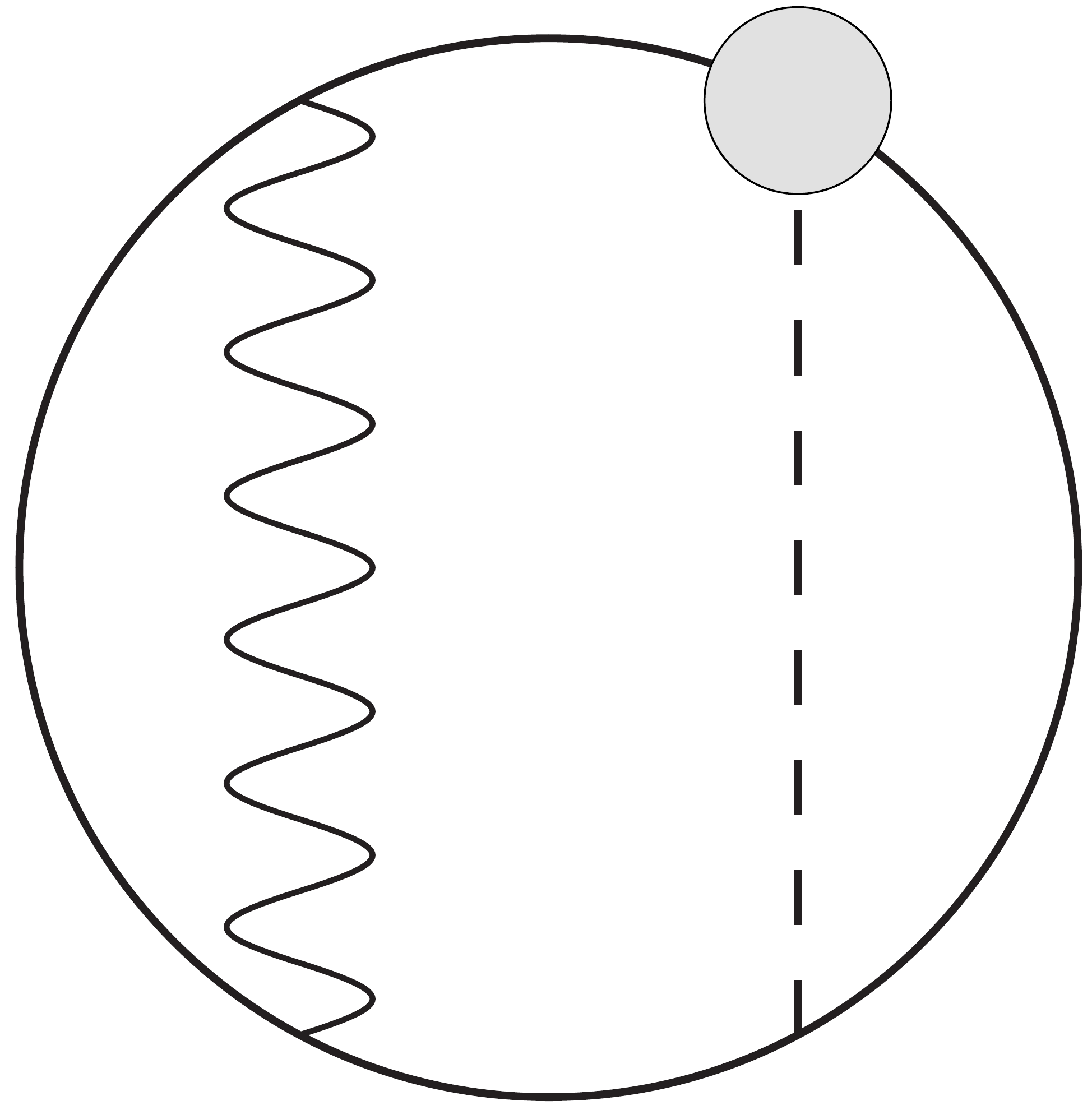}}}~,\nonumber\\
\partial_I \vcenter{\hbox{\includegraphics[width=1.5cm]{figs/a311-eps-converted-to.pdf}}} = \vcenter{\hbox{\includegraphics[width=1.5cm]{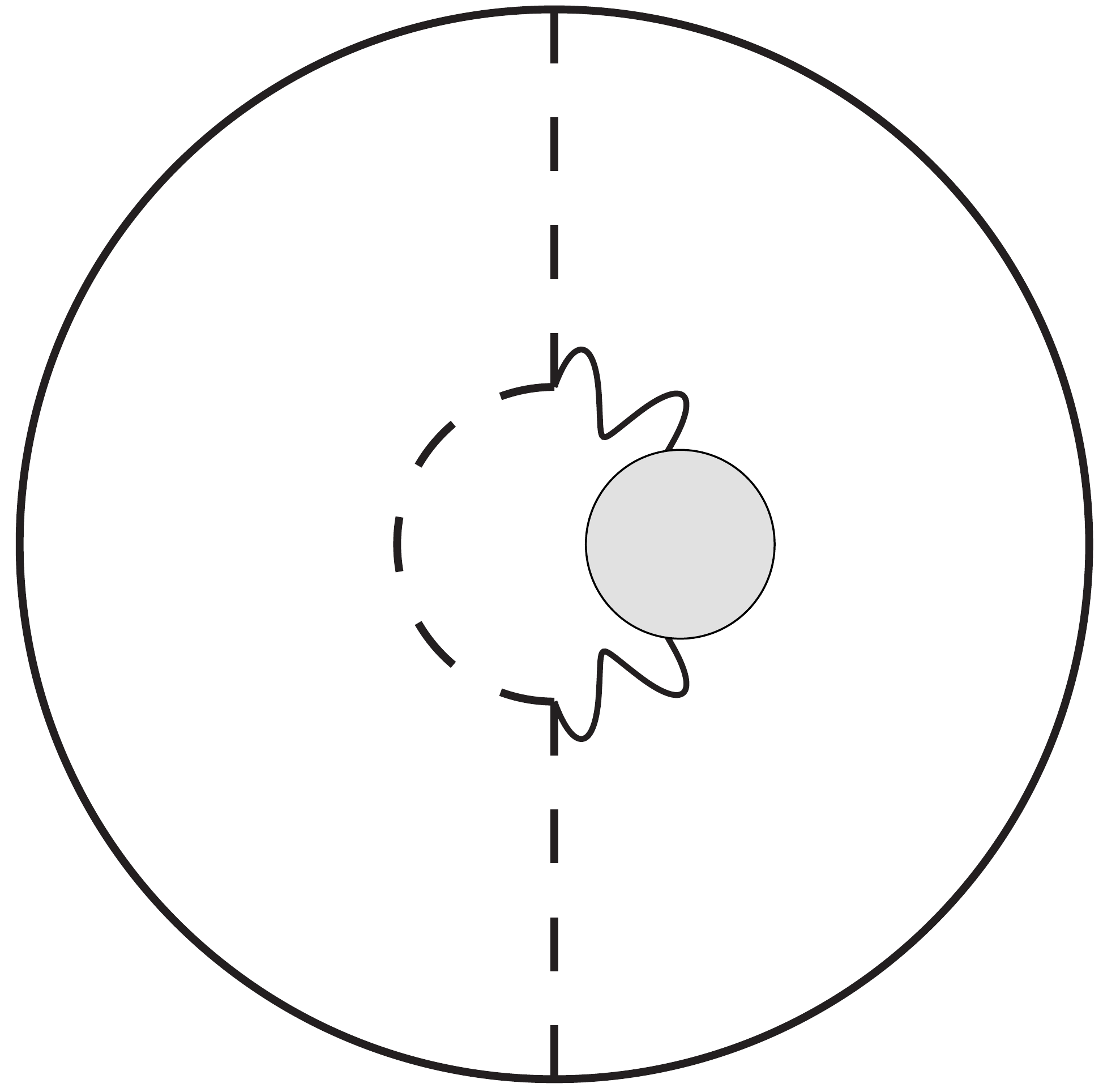}}} + 2\,\vcenter{\hbox{\includegraphics[width=1.5cm]{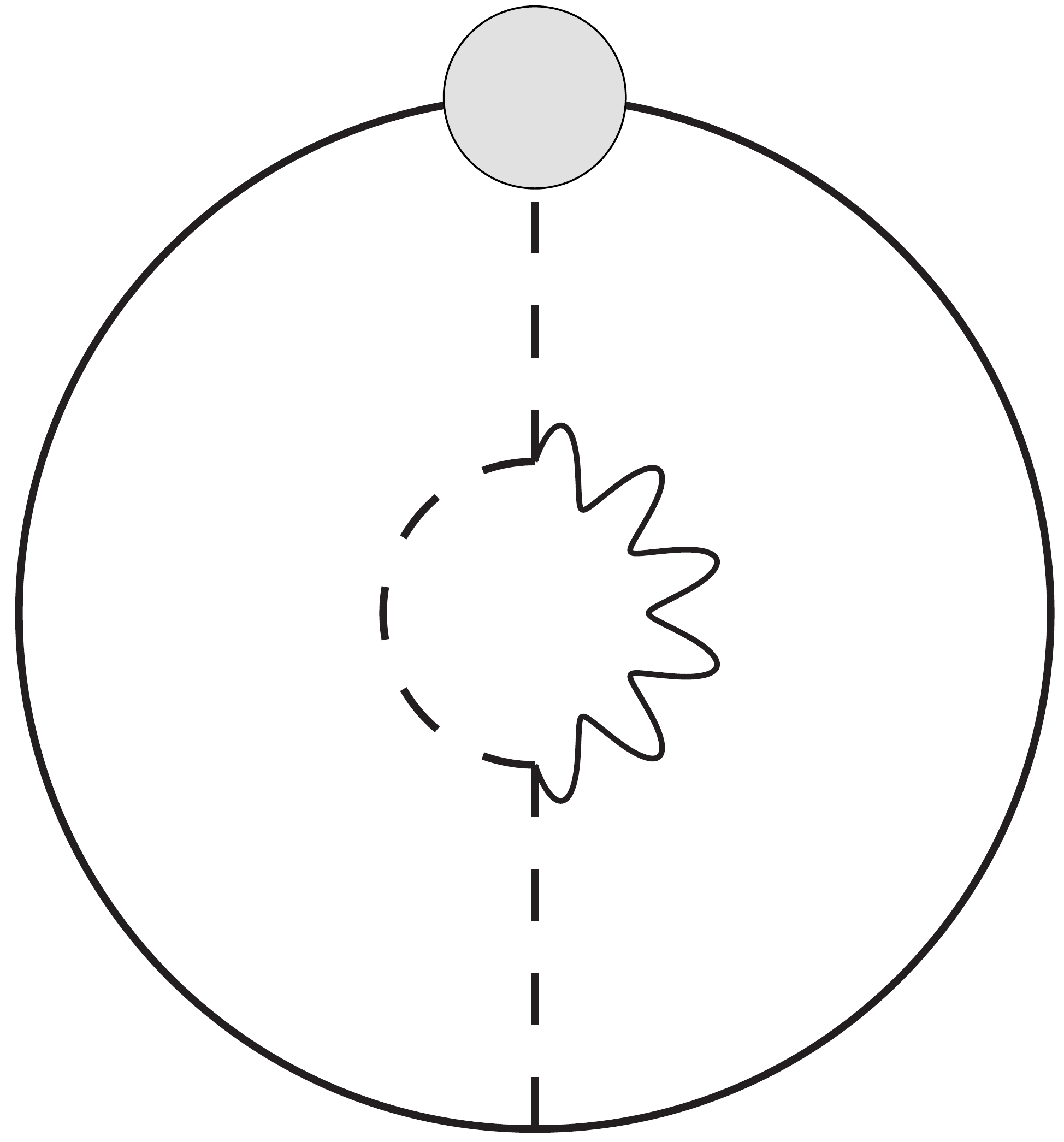}}}~.
\end{align}
Thus leading to
\begin{align}
\partial_I\tilde{A} \supset a_{10}^{(3l)} \left(\vcenter{\hbox{\includegraphics[width=1.3cm]{figs/a310deriv1-eps-converted-to.pdf}}} + 2\,\vcenter{\hbox{\includegraphics[width=1.3cm]{figs/a310deriv2-eps-converted-to.pdf}}}\right) + a_{11}^{(3l)} \left(\vcenter{\hbox{\includegraphics[width=1.3cm]{figs/a311deriv1-eps-converted-to.pdf}}} + 2\,\vcenter{\hbox{\includegraphics[width=1.3cm]{figs/a311deriv2-eps-converted-to.pdf}}}\right)~.
\end{align}

Similarly, the two objects on the r.h.s of the equation, $T_{IJ}$ and $\beta^J$, consist of tensor structures with two and one open index, respectively. The $\beta^J$ are the gauge, Yukawa and quartic $\beta$-functions
we are interested in. The tensor structures relevant for the two structures in $A$ shown above are given by
\begin{align}
T_{IJ} \supset t_1^{(1l)} \vcenter{\hbox{\includegraphics[width=1.5cm]{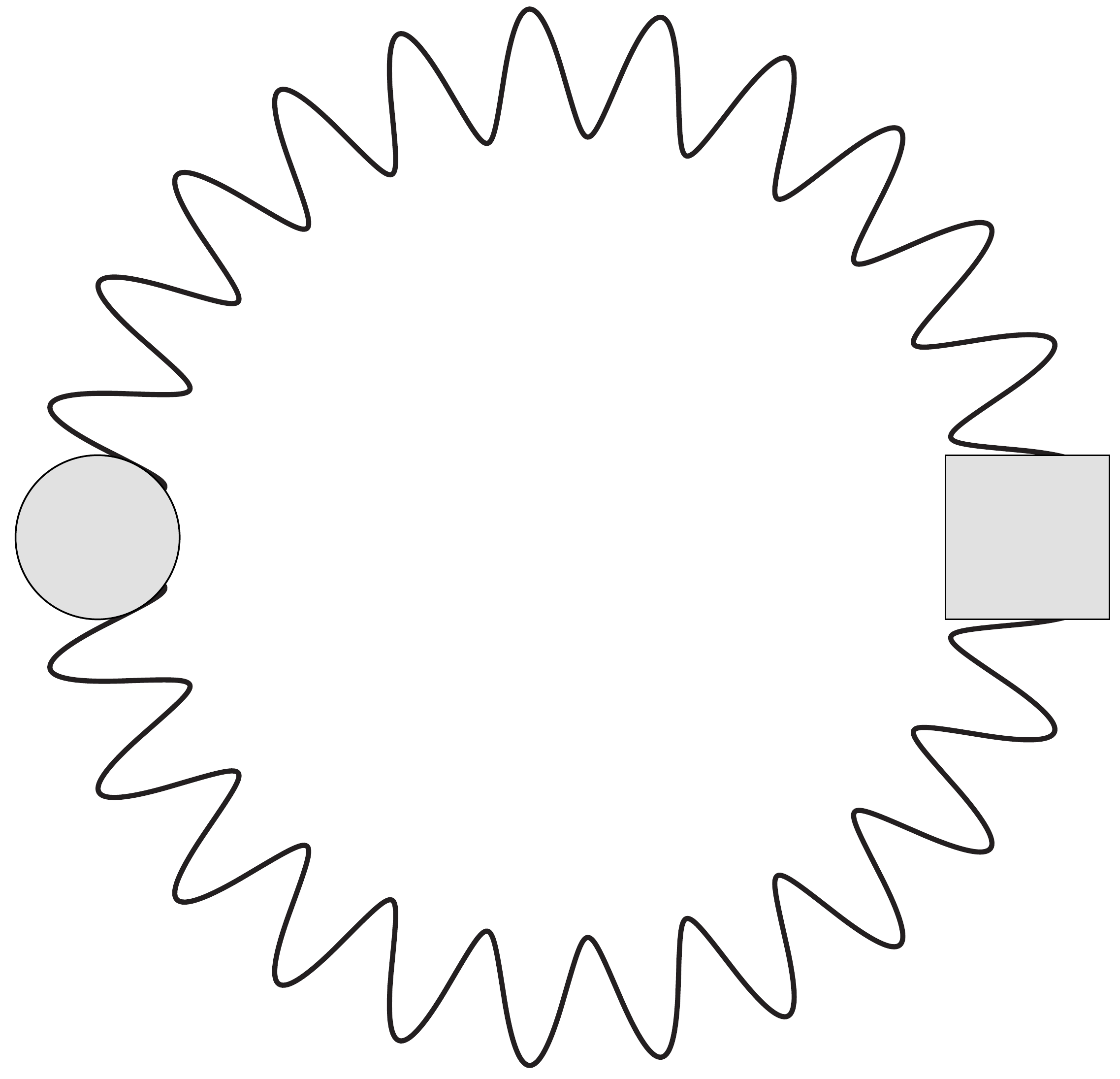}}} + t_{4}^{(2l)} \vcenter{\hbox{\includegraphics[width=1.5cm]{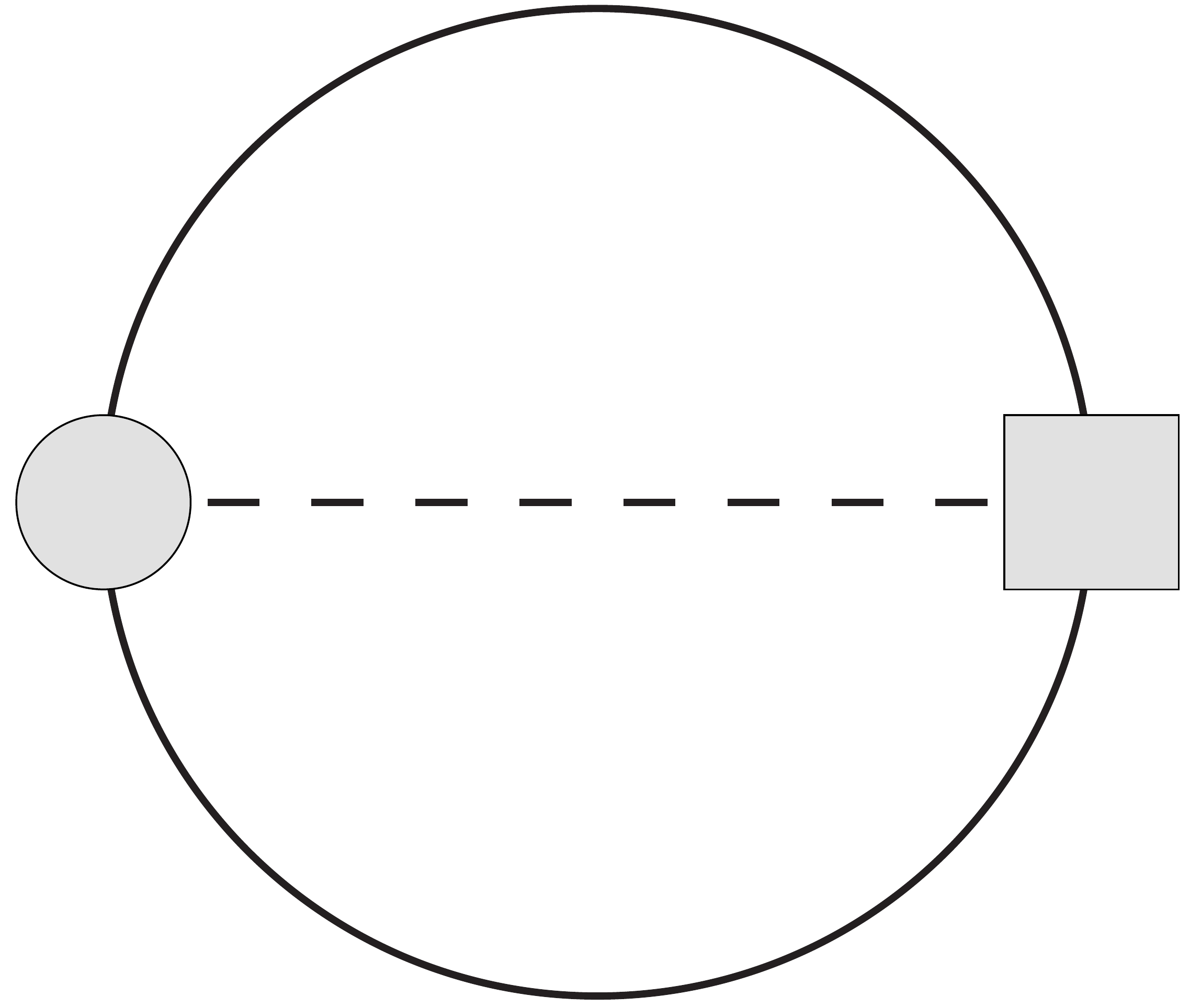}}}
\end{align}
and
\begin{align}
\beta^J \supset g_6^{(2l)} \vcenter{\hbox{\includegraphics[width=1.8cm]{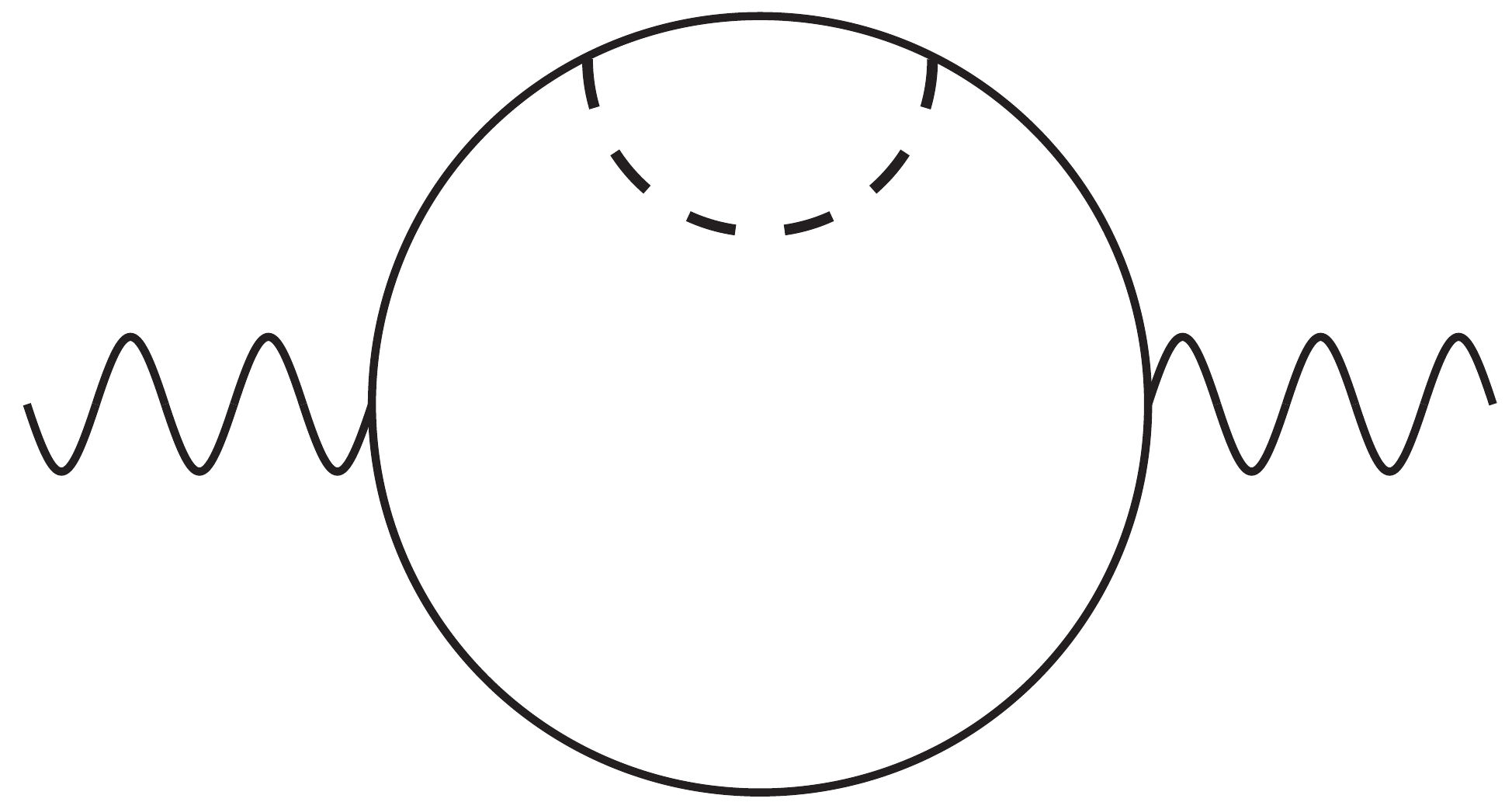}}} + g_{7}^{(2l)} \vcenter{\hbox{\includegraphics[width=1.8cm]{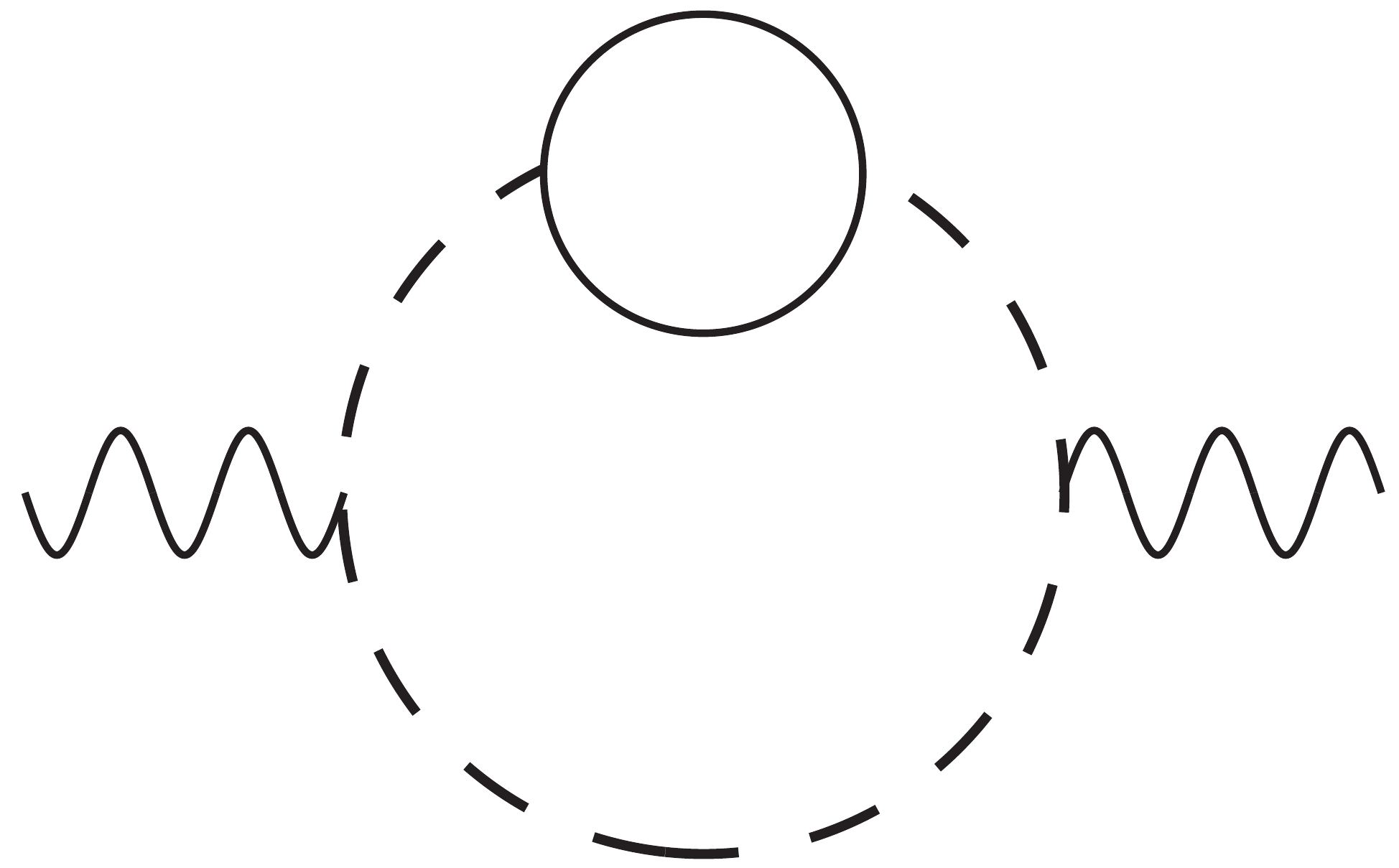}}}\\ + y_1^{(1l)} \vcenter{\hbox{\includegraphics[width=1.8cm]{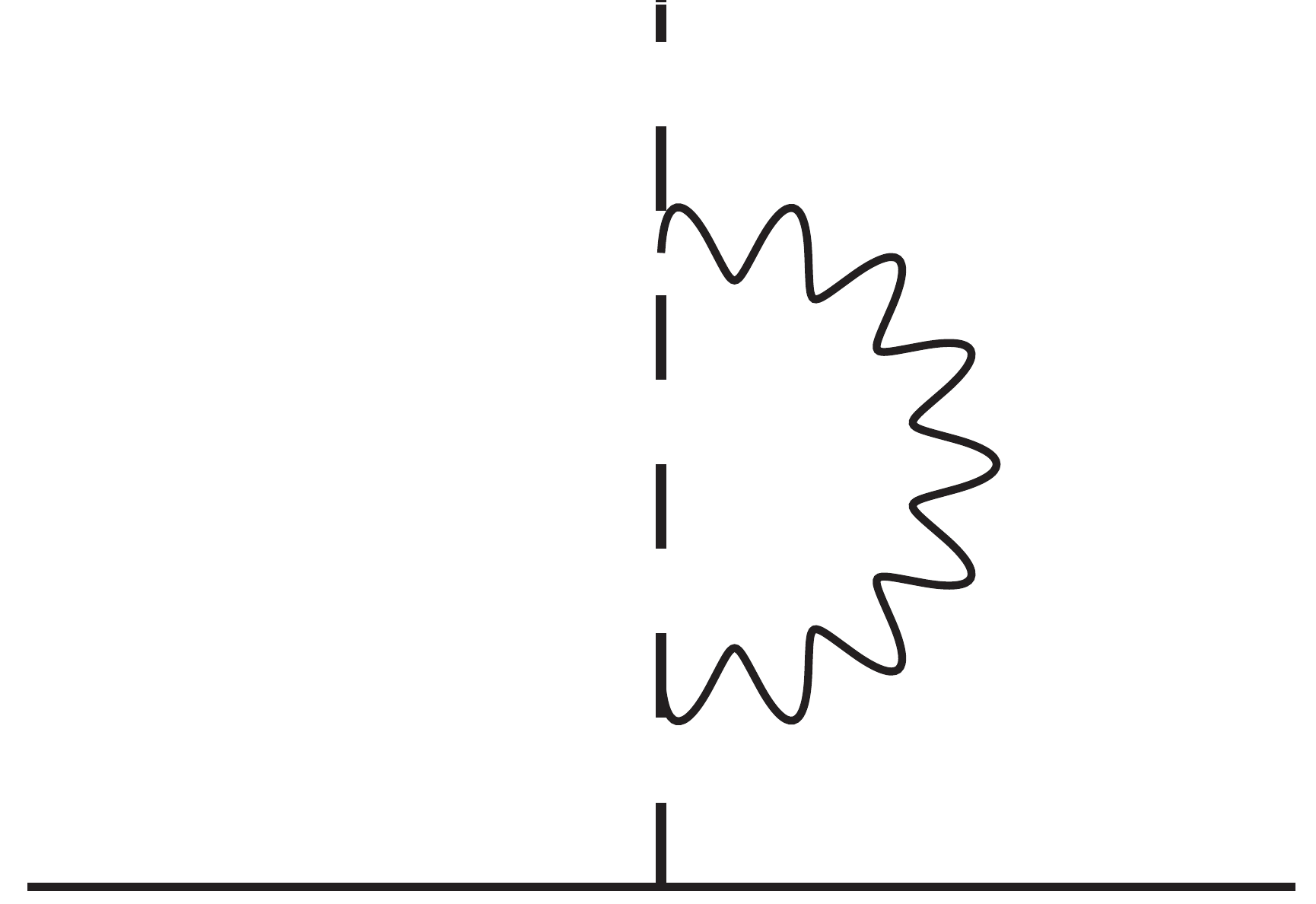}}} + y_{2}^{(1l)} \vcenter{\hbox{\includegraphics[width=1.8cm]{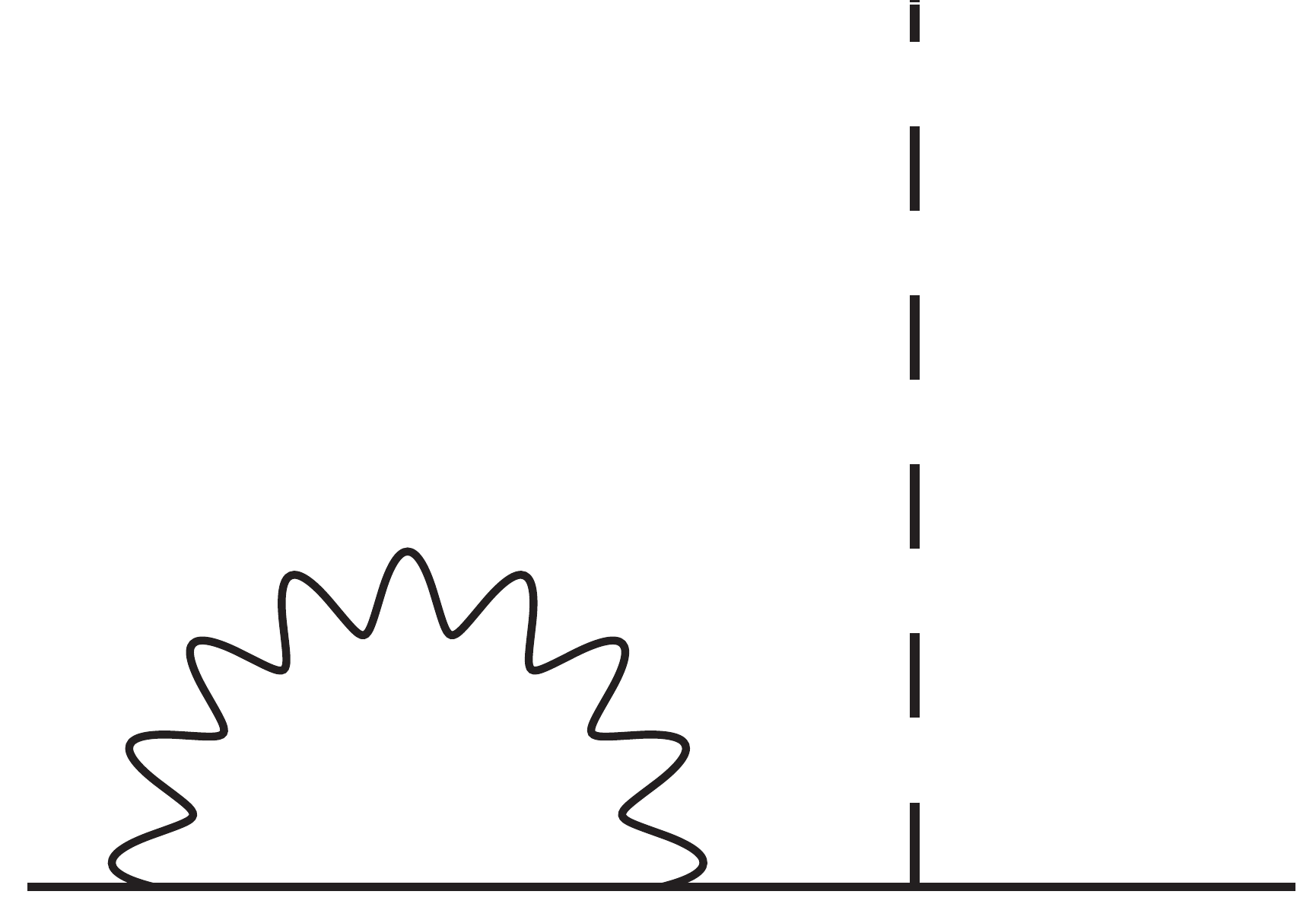}}}~.
\end{align}
Identifying tensor structures on the two sides of Osborn's equation, we obtain 4 equations for the coefficients:
\begin{align}
a^{(3l)}_{10} = t^{(1l)}_1 g^{(2l)}_6~,\qquad a^{(3l)}_{11} &= t^{(1l)}_1 g^{(2l)}_7~,\nonumber\\
2 a^{(3l)}_{10} = t^{(2l)}_4 y^{(1l)}_1~,\qquad 2 a^{(3l)}_{11} &= t^{(2l)}_4 y^{(1l)}_2~,
\end{align}
which can be solved for
\begin{align}
g^{(2l)}_7 y^{(1l)}_2 = g^{(2l)}_6 y^{(1l)}_1~.
\end{align}
Thus, Osborn's equation connects coefficients of two-loop gauge coupling $\beta$-function with coefficients in the one-loop Yukawa $\beta$-function.
Going to higher orders, the equations connect $L$-loop gauge to $(L-1)$-loop Yukawa and $(L-2)$-loop quartic $\beta$-function coefficients.
This feature can be used in two ways; Either as a cross-check of explicit calculations or as a way to determine additional coefficients in the general result from already-known ones.

Starting from three loops for the Yukawa $\beta$-function and four loops for the gauge $\beta$-function, non-trivial contributions from fermionic loops with an odd number of $\gamma_5$ arise in chiral theories like the SM.
Sample tensor structures are given by:
\begin{center}
\includegraphics[width=8.0cm]{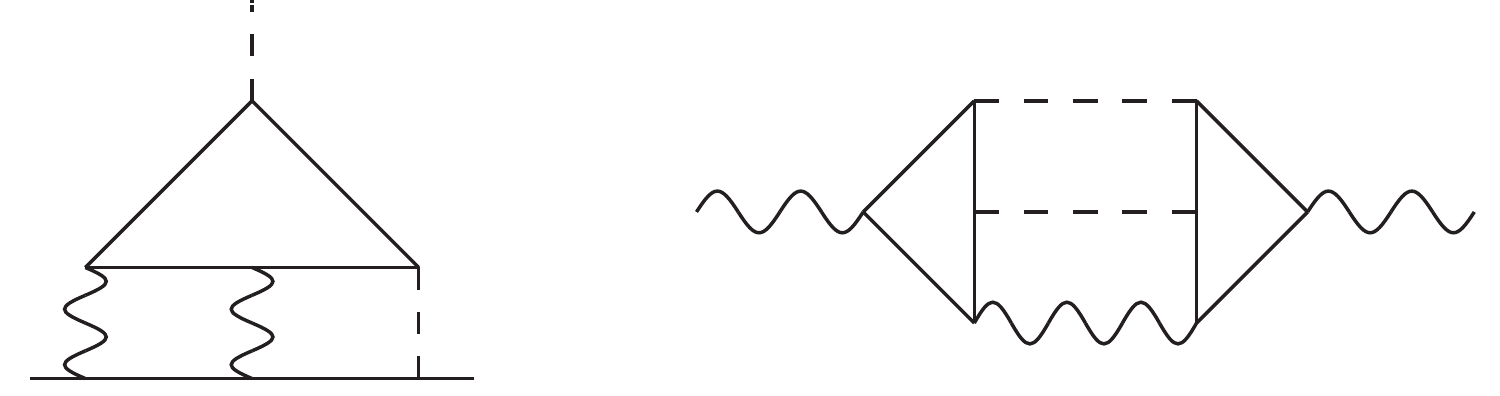}
\end{center}
While the three-loop Yukawa $\beta$-functions can be computed by treating $\gamma_5$-odd traces as in four dimensions, this is not the case for the four-loop gauge $\beta$-functions.
Here, sub-divergences appear in the relevant Feynman diagrams and thus a $D$-dimensional treatment would be necessary.
However, Osborn's equation provides us with a way to resolve this issue without the need to actually solve the "$\gamma_5$-problem".
Observe, that the two sample tensor structures above arise from the derivative of
\begin{center}
\includegraphics[width=2.0cm]{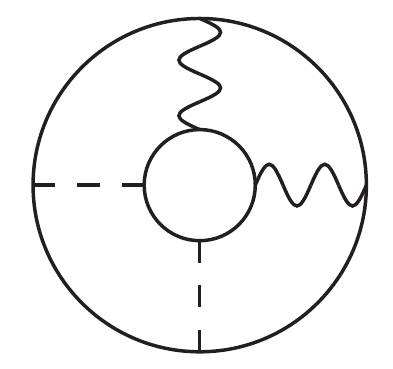}
\end{center}
w.r.t.~gauge or Yukawa couplings. Thus their coefficients are related. While no guarantee exists that this is the case for all $\gamma_5$-odd coefficients, at least at the 4--3--2-loop order
all $\gamma_5$-odd gauge $\beta$-function coefficients are related, miraculously, to the $\gamma_5$-odd Yukawa $\beta$-function coefficients \cite{Poole:2019txl}.

With the $\gamma_5$-odd $\beta$-function coefficients fixed, the respective tensor structures can be evaluated for the SM.
The $\gamma_5$-even contributions can be calculated by dropping traces with an
odd number of $\gamma_5$. These contributions have been computed in Ref.~\cite{Davies:2019onf} and combined with the $\gamma_5$-odd
contributions, finalizing the first computation of four-loop $\beta$-functions in a chiral gauge theory.

As a complete parametrization of the gauge--Yukawa--scalar $\beta$-functions at 4--3--2-loop order is available,
results in specific models can be used to fix the free coefficients. Combining the four-loop gauge coupling $\beta$-functions in the SM, the three-loop Yukawa
$\beta$-functions in the SM and 2HDMs \cite{Bednyakov:2014pia,Herren:2017uxn} with the WCCs fixes the vast majority of coefficients. However, some coefficients remain unconstrained and thus
additional models have to be considered. Two independent calculations using different toy models, Refs. \cite{Bednyakov:2021qxa} and \cite{Davies:2021mnc}, have recently been completed, fixing the remaining coefficients
and, as a consequence, extending the general results to 4--3--2-loop order.

Our results have been implemented in the program \texttt{RGBeta} \cite{Thomsen:2021ncy}, which allows the user to specify a model and returns the respective $\beta$-functions.
The $\beta$-functions can easily be implemented in \texttt{C++} programs using the template library \texttt{RGE++} \cite{Deppisch:2020aoj}.

\section{Ambiguities in Yukawa $\beta$-functions}
Starting from three loops, Yukawa and scalar $\beta$-functions are ambiguous and not necessarily finite anymore. This was first discovered in the context
of the Yukawa matrix $\beta$-functions in the SM \cite{Bednyakov:2014pia} and 2HDMs \cite{Herren:2017uxn}.

The appearance of ambiguities can be traced back to the need to take square roots of complex matrices.
They are necessary to compute fermion and scalar wave-function renormalization constants. In the fermionic case, these are given by
\begin{align}
Z_f = 1 - K_\epsilon\left(\sqrt{Z_f}^\dagger \Sigma(Q^2) \sqrt{Z_f}\right)~,
\end{align}
where the operator $K_\epsilon$ takes the divergent part and $\Sigma(Q^2)$ is the fermion self-energy.
Square roots of matrices can be defined in multiple, equivalent ways:
\begin{align}
Z_f = \sqrt{Z_f}^\dagger \sqrt{Z_f} = \left(U\hat{\sqrt{Z_f}}\right)^\dagger U\hat{\sqrt{Z_f}}~.
\end{align}
Here $U$ is a unitary matrix, possibly containing poles in $\epsilon$. Different choices of $U$ lead to the same $Z_f$, but different anomalous dimensions:
\begin{align}
\gamma_f &= \sqrt{Z_f}^{-1}\mu\frac{\mathrm{d}}{\mathrm{d}\mu}\sqrt{Z_f} = U^\dagger\hat{\sqrt{Z_f}}^{-1}\left(\mu\frac{\mathrm{d}}{\mathrm{d}\mu}\hat{\sqrt{Z_f}}\right)U + U^\dagger\mu\frac{\mathrm{d}}{\mathrm{d}\mu} U~.
\end{align}
As the anomalous dimensions of fermion fields  enter Yukawa $\beta$-functions these also depend on the choice of $U$.
Simply choosing $U = \mathbb{1}$ leads to Hermitian square roots, but divergent Yukawa matrix $\beta$-functions.

It was proven, that the divergences in the anomalous dimensions are tightly connected to global symmetries of the kinetic terms of fermions and scalars
and are the sole source of divergences in the $\beta$-functions \cite{Herren:2021yur}:
\begin{theo}{RG-finiteness}
	The divergent part of \emph{any} set of MS/$\overline{\mathrm{MS}}$ RG functions $ (\beta_I,\, \gamma) $ satisfy 
		\begin{equation*} 
		\hgamma^{(n)} \in \mathfrak{g}_F \andeq \hbeta^{(n)}_I = - \big( \hgamma^{(n)}\, g \big)_I \,, \qquad n \geq 1\,.
		\end{equation*}
	This property of the RG functions is referred to as RG-finiteness.
\end{theo}
Here the superscript $(n)$ denotes the term proportional to $\epsilon^{-n}$ and $\mathfrak{g}_F$ is the Lie algebra associated to the global symmetry group of the kinetic terms.

Working with ambiguous (or even worse, divergent) $\beta$-functions is, at best, inconvenient, thus having an unique prescription for a finite $\beta$-function is desirable.
The framework of the local RG offers a solution to this, the so-called $B$-function, related to the conventional $\beta$-function by \cite{Fortin:2012hn}
\begin{align}
B_I = \beta_{I} - (\upsilon\, g)_I~.
\end{align}
$B$ is unique, while $\upsilon$ and $\beta$ depend on the choice of square roots. In Ref.~\cite{Herren:2021yur} $\upsilon$ was computed for the choice of
Hermitian square roots at three loops and the results presented in Refs.~\cite{Bednyakov:2021qxa,Davies:2021mnc} are compatible with this choice.

\section{Conclusion and Outlook}
Recently the general gauge coupling and Yukawa matrix $\beta$-functions have been computed at four and three loops, respectively. These results rely on Osborn's equation for fixing the
non-trivial $\gamma_5$ contributions to the gauge coupling $\beta$-function, as well as on the proper understanding of ambiguities and divergences in Yukawa matrix $\beta$-function.

With the $\beta$-functions available at  4--3--2-loop order, the next step is to derive a basis of tensor structures and the relations between the coefficients at 5--4--3-loop order.
This would allow the determination of the three-loop quartic $\beta$-function and the investigation of the $\gamma_5$-odd contributions at this order. Should the \emph{miracle} repeat itself, a computation
of the general $\beta$-functions at 5--4--3-loop order is, in principle, feasible. Furthermore, it would be interesting to study if Osborn's equation imposes constraints in
pure gauge theories at high loop orders.

\section*{Acknowledgements}
We thank Joshua Davies and Anders Eller Thomsen for collaboration on the discussed topics and for carefully reading the manuscript.

\paragraph{Funding information}
FH acknowledges support by the Alexander von Humboldt foundation. This document was prepared using the resources of the Fermi National Accelerator Laboratory (Fermilab), a U.S. Department of Energy, Office of Science,
HEP User Facility. Fermilab is managed by Fermi Research Alliance, LLC (FRA), acting under Contract No. DE-AC02-07CH11359.

\bibliography{radcor_herren.bib}

\nolinenumbers

\end{document}